# Adaptive hedging horizon and hedging performance estimation

Wang Haoyu[a], Junpeng Di[b], Qing Han[b]

a. Xiamen University, School of Management, Xiamen
b. Shanghai Academy of Social Sciences, Institutes of Economics, Shanghai

**Abstract**

In this study, we constitute an adaptive hedging method based on empirical mode decomposition (EMD) method to extract the adaptive hedging horizon and build a time series cross-validation method for robust hedging performance estimation. Basing on the variance reduction criterion and the value-at-risk (VaR) criterion, we find that the estimation of in-sample hedging performance is inconsistent with that of the out-sample hedging performance. The EMD hedging method family exhibits superior performance on the VaR criterion compared with the minimum variance hedging method. The matching degree of the spot and futures contracts at the specific time scale is the key determinant of the hedging performance in the corresponding hedging horizon.

## 1. Introduction

Hedging performance is the central concern of hedgers. Among the determinants of the hedging performance, hedging horizon is the main factor. Geppert (1995) and Chen et al. (2004) pioneer the research on this topic, and the latter especially point out the relationship between the frequency of contract time series and hedging horizon. Lien and Shrestha (2007) extend this idea by introducing the wavelet method to decompose the contract time series into different time scales corresponding to the hedging horizon. They find that compared with conventional methods, the wavelet hedging method does not show superior performance based on the minimum variance criterion, and, thus, the wavelet hedging method is mainly used to solve the sample reduction problem of static hedging strategy.

Some studies extend the usage of the wavelet method (Conlon and Cotter, 2012; Sultan et al., 2019).

The wavelet method decomposes the contract time series into the specific frequency time series, which is usually the multiples of 2. This artificial prior setting ignores and deteriorates the intrinsic frequency structure of the contract time series, behind which is the trader behavior and economic fundamental (Fryzlewicz et al., 2006; 2013). Extract the intrinsic frequency structure of the contracts directly is more intuitive and rational. Doing so does not distort the information contained in the contracts, and, thus, better hedging performance appears with the less distortion. Basing on this idea, this study introduces an adaptive filtering method, empirical mode decomposition (EMD), which can extract the intrinsic frequency structure of the contract time series. Adaptive hedging horizons are defined as the cycles of this intrinsic frequency structure. Using the EMD method, we analyze the matching degree of the spot and futures contracts, which is tested as the main determinant of the hedging performance. We find the same results with Lien and Shrestha (2007), that is, time frequency methods are not superior to conventional methods for determining the hedging performance on the variance reduction criterion. However, we find that the EMD-based hedging methods are superior to the conventional methods on estimating the hedging performance of the value-at-risk (VaR) criterion.

Hedging performance estimation, especially the out-sample performance estimation, is a core topic, but few studies provide comprehensive discussion (Cotter and Hanley, 2006; Lien, 2006, 2010, 2012; Wang et al., 2015). In most literature, the out-sample performance estimation is through randomly splitting the contracts time series into training and testing intervals, calculating the hedge ratio in the training interval, and estimating the hedge ratio's performance in the testing interval (Salvador and Arago, 2013; Xu and Lien, 2020; Lai, 2021). This customary method has stochastic, even data snooping prior settings of splitting sample, which consequently cause non-robustness and sample limit problem because the history price time series is only one realization of many stochastic routes. Based on the cross-validation method in machine learning, this study constitutes a time series cross-validation method for hedging performance estimation, which can solve the stochastic splitting sample problem and promote the robustness of the estimation. Through this method, we confirm that the time frequency hedging method in Lien and Shrestha (2007) does not show superior performance on the variance reduction criterion. We also analyze the method performance on the VaR criterion, where we observe the superior performance of the EMD method family.

This study has four main contributions to the literature. First, it constitutes an adaptive hedging method, which can generate the adaptive hedging horizon by extracting the contract frequency structure. Second, it constitutes a time series cross-validation method for hedging performance estimation, which can avoid the estimation bias of arbitrary split of the training and testing time series. Third, it finds the inconformity of the in-sample and out-sample hedging performance estimation and extends the time frequency method usage by Lien and Shrestha (2007). Fourth, it finds that the key determinant of the hedging performance is the matching degree of the spot and futures contracts in the corresponding adaptive hedging horizon.

The rest of the study proceeds as follows. Section 2 introduces the methodology in this study. Section 3 presents the empirical results of in-sample and out-sample performance. Section 4 draws the conclusion.

## 2. Methodology

This section presents the conventional hedging method and then introduces one of the time frequency hedging methods, the wavelet hedging method (Lien and Shrestha, 2007), which provides the basic pattern of the time frequency methods. Subsequently, we constitute the EMD hedging method that can extract the frequency structure of the contracts, which we define as the adaptive hedging horizons. We constitute the time series cross-validation estimation for hedging performance and path performance based on the cross-validation method in machine learning and conventional hedging performance estimation.

### 2.1. Conventional method

The conventional methods are the minimum hedging method and its derivatives, including the error correction method (Chen, 2004) and extended error correction method (2005).

#### 2.1.1. Minimum variance hedging method

The hedging portfolio is constructed by taking $C_S$ units of a long spot position and $C_F$

units of a short futures position, and the hedging horizon is from time 0 to time $T$. The value of hedging portfolio is

$$V_{H,t} = C_S S_t - C_F F_t,$$

where $S_t$ and $F_t$ is the price of the spot and the futures at time t, respectively. The change of the hedging portfolio value over the hedging period is

$$\Delta V_H = C_S \Delta S_T - C_F \Delta F_T,$$

where $\Delta S_T = S_T - S_0$ and $\Delta F_T = F_T - F_0$.

The aim of the minimum variance hedging is to minimize the variance of $\Delta V_H$, and the hedge ratio is

$$h = \frac{cov(\Delta S, \Delta F)}{var(\Delta F)}.$$

This hedge ratio is obtained by solving the following function:

$$\min_h var(\Delta S - h\Delta F),$$

where the $h = C_F/C_S$.

In empirical cases, the optimal hedge ratio is estimated through the regression

$$\Delta S = a + \beta \Delta F + e,$$

where the parameter $\beta$ is the estimated optimal hedge ratio $h$.

The optimal hedge ratio is driven by covariance of the change of the spot price and the futures price, the variance of the futures price change, and the hedging horizon. Among which, only the hedging horizon is determined by hedgers and empirically has great influence on the hedging ratio.

### 2.1.2. ECM hedging method

Chen et al. (2004) analyze the hedging horizon effect on the hedge ratio and hedging performance. They find significant difference of hedging performance between short and long horizons. In the long horizon, the hedge ratio tends to converge to naïve hedging ratio that equals to 1, whereas the hedge ratio is usually below the naïve hedge ratio in the short horizon. To consider these hedge ratios together, they constitute an error correction model,

$$\Delta S_t = a + \beta \Delta F_t + \alpha_S S_{t-1} + \alpha_F F_{t-1} + e_t,$$

where $S_{t-1}$ and $F_{t-1}$ are the lag spot price and futures price. The intuitive of this model is introducing the long-term convergence of spot price and futures price following the cointegrating equation,

$$S_t = a + bF_t + u_t,$$

into the short-term regression model.

### 2.1.3. Extended ECM hedging method

Lien and Shrestha (2005) extend Chen's model by adding lags of $\Delta S_t$ and $\Delta F_t$ to improve the fitness, which takes the risk of overfitting. The number of lags is determined by the model's Akaike information criterion,

$$\Delta S_t = a + \beta_1 \Delta F_t + \beta_2 u_{t-1} + \sum_{i=1}^{m} \gamma_{si} \Delta S_{t-i} + \sum_{j=1}^{n} \gamma_{fj} \Delta F_{t-j} + e_t,$$

where $u_t$ is the residual from the cointegrating regression,

$$S_t = a + bF_t + u_t.$$

## 2.2. Time frequency hedging method

Time frequency hedging methods are derived from the idea that contracts have certain characteristics at the specific time scales and that the hedgers calculate the hedge ratio using the characteristics of time scales corresponding to their hedge horizon. The wavelet hedging method assume the time scales are the multiple of 2, which distorts the information in the contracts. By contrast, the EMD method extracts the characteristic time scales adaptively, which we define as the adaptive hedging horizon. In addition, the wavelet method generally extracts the fuzzy characteristics, whereas the EMD method extracts the finer characteristics (Huang et al., 2003).

### 2.2.1. Time dimension method: wavelet method

Wavelet analysis is a well-developed time-frequency filtering method is first used in engineering and physics. However, since the filtering method by Hodrick and Prescott (1997), filtering methods, especially time frequency filtering method, are increasingly becoming popular (Freylewicz et al., 2006; 2013). The wavelet method is one of the most popular time frequency filtering methods and is applied by some studies (Lien and Shrestha, 2007; Conlon and Cotter 2012; Sultan et al., 2016) on hedging analysis.

The basic idea of the wavelet method on hedging analysis is decomposing the original spot and futures return series into different time scales and calculating the hedge ratio of each time scales. In most studies, the time scale lengths are $2^i, i = 1,2,...,n$, and each scale contains its own characteristic corresponding to its length. Facing different horizons, hedgers only need to choose the time scale approximate to the hedging horizon and take the hedge ratio at that scale as the optimal hedge ratio at that horizon. This method does not need to differentiate the sample to match the hedging horizon as conventional methods do; thus, it avoids the sample reduction problem of the long hedging horizon. However, its performance on short hedging horizon has no advantages compared with that of conventional methods (Lien and Shrestha, 2007).

We give a brief mathematical description of wavelet method. First, determine the mother function and the father function, which are parameters distinguishing the wavelet method. Then, decompose the spot return and futures return.

$$\Delta S_t = B_{J,t}^S + D_{J,t}^S + D_{J-1,t}^S + \cdots + D_{1,t}^S,$$
$$\Delta F_t = B_{J,t}^F + D_{J,t}^F + D_{J-1,t}^F + \cdots + D_{1,t}^F.$$

Finally, estimate $J$ regressions using the $J$ decompositions.

$$D_{j,t}^S = \theta_{j,0} + \theta_{j,1} D_{j,t}^F + e_{j,t}$$

The minimum variance hedge ratio of the $j$th time scale is given by the estimate of $\theta_{j,1}$. From the description above, we can see that the wavelet method can be integrated with conventional methods, like minimum variance (MV) hedge and ECM hedge, by using the conventional methods on the decomposed time series (Sultan et al., 2019).

### 2.2.2. Time dimension method: EMD

EMD is an adaptive time frequency method designed for non-stationary and non-linear time series (Huang et al., 1998). Huang et al. (2003) extend it to financial time series analysis and discuss its advantages on analyzing financial data over Fourier transform and the wavelet method. As pointed by Chen et al. (2004), the return series of spot and futures are not certainly stationary and linear. Even worse, in some cases, the spot return is stationary, but futures return is non-stationary. EMD can solve this problem, which is a marginal advantage of this method.

The main advantages of the EMD method are its adaptiveness and simplicity. Regarding adaptiveness, the EMD method filters the time series by the characteristics of time series per se, which need no basis functions such as Fourier transform and wavelet method. Without prior setting, this method requires minimum artificial set up, making it possible to extract the characteristics of the time series purer than other filtering methods. Regarding simplicity, the EMD method has concise algorithms that require no heavy computation, which is the main advantage for wide application. The algorithm of the EMD method is to get the characteristic time series, named intrinsic mode function (IMF), which satisfies the following two conditions:

(1) In the whole data set, the number of extrema and the number of zeros-crossings must either equal or differ at most by 1.
(2) At any point, the mean value of envelope defined by the local maxima and the envelope defined by the local minima is 0.

To get the defined IMF, the procedures are as follows:

(1) Identify all the local extrema of the original time series $X(t)$ and then connect all the local maxima by a cubic splice as the upper envelope. Repeat the procedure for the local minima to produce the lower envelope. The mean of the upper and lower envelopes is then designated as $m_1$.

(2) Take the difference of the $X(t)$ and the $m_1$ and the differenced time series as $h_1$. The $h_1$ is taken as the original time-series and sifted several times $k$ following step 1, until we obtain the first time series that satisfies the two conditions and take it as IMF, $c_1$. Separate the $c_1$ from the rest of the data by

$$X(t) - c_1 = r_1.$$

(3) $r_1$ is treated as the new time-series and subjected to the same sifting process in step 2. This procedure can be repeated to obtain all the subsequent $r_j$'s, and the original time-series is decomposed as

$$X(t) = \sum_{j=1}^{n} c_j + r_n,$$

where the residue $r_n$ is a constant, a monotonic function, or a function with only one maximum and one minimum form that no more IMFs can be extracted.

The IMFs contain the characteristics of the original time series at each adaptive time scale because the EMD method sets no prior scales like wavelet method. Setting the hedging horizon consistent with the contract's intrinsic scales is a rational idea, behind which are the behavior of the traders and economic fundamentals that drive the dynamics of the contract. However, a hedger who has an aimed hedging horizon should focus on the behavior of the scales consistent with his hedging horizon as the longer scales are not deterministic of the hedging performance.

The EMD hedging method has similar ideas with the wavelet hedging method because the two methods focus on the time series behavior at different scales and extract the related characteristics to determine the hedge ratio. The basic difference between the EMD hedging method and wavelet hedging method is that the former sets no prior scales because the predetermined sales can distort the result if the behavior of the traders and the economic fundamentals are not consistent with the predetermined scales.

Based on the above analysis, this study constitutes the EMD hedging method family, including the vanilla EMD hedging method, the sample saving EMD hedging method, and the aggregate EMD hedging method.

The vanilla EMD hedge is the simplest hedging method using EMD. The idea is simple, that is, matching the spot and futures time series characteristics in the certain hedging horizon, so the other scales, particularly the larger scales, of the time series' characteristics need no consideration because they are not matched with the hedging horizon. The vanilla EMD hedge ratio of certain hedging horizon can be calculated through the OLS regression below,

$$\Delta IMF_{i,t}^S = a + \beta \Delta IMF_{i,t}^F + e_{i,t},$$

where $\Delta IMF_{i,t}^S$ is the difference series of $ith$ spot IMF, $\Delta IMF_{i,t}^F$ is the difference series of $ith$ futures IMF, and $\beta$ is the hedge ratio of the corresponding hedging horizon. Some details in this equation are supposed to be pointed out. First, the difference operation corresponds to the hedging horizon. For example, if the hedging horizon is 3 days, then the difference operation $\Delta$ for $IMF$ is $IMF_{i,t} - IMF_{i,t-3}$, assuming the time unit is day. Second, the differenced time series is the time series per se, not the manipulated time series, like log return time series. This is because the manipulation distorts the characteristics and information of the original time series.

The main advantage of the time dimension method is solving the sample reduction problem of static hedging model (Lien and Shrestha, 2007). The EMD hedging method has this same advantage as the wavelet hedging method because they dig the time series characteristics of the certain hedging horizon. Like the wavelet method, the sample saving EMD hedging method is

$$IMF_{i,t}^S = a + \beta IMF_{i,t}^F + e_{i,t},$$

where $IMF_{i,t}^S$ is the $ith$ spot IMF, $IMF_{i,t}^F$ is the $ith$ futures IMF, and $\beta$ is the hedge ratio of the corresponding hedging horizon.

As the EMD method extracts the characteristics of the time series at different time scales, the scales smaller than hedging horizon naturally contain relevant characteristics that can bring more accurate estimation of the hedge ratio. Thus, we constitute the aggregate EMD hedging method for this consideration, and the regression equation is

$$\sum_{cycle_i \leq horizon} IMF_{i,t}^S = a + \beta \sum_{cycle_i \leq horizon} IMF_{i,t}^F + e_{i,t},$$

where $cycle_i$ is the cycle of $ith$ IMF, $horizon$ is the hedging horizon, $\beta$ is the hedge ratio, and the other notations are the same with those in the former equation.

## 2.3. Performance models

### 2.3.1. Conventional performance models

The performance models are aimed to measure the out-sample hedging performance. Based on the concept of hedging, the variance reduction criterion measures the performance of the hedging methods on the degree of decreasing the hedge portfolio variance, which is given by

$$HE_{variance} = 1 - \frac{var(\Delta V_H)}{var(\Delta S)}$$

where $var(\Delta V_H)$ and $var(\Delta S)$ are variance of the returns for the hedging portfolio and the spot contracts, respectively.

Besides the variance reduction criterion, we study the higher moment of the hedging performance by using the VaR criterion, which estimates the maximum portfolio expected loss for a given confidence level $\alpha$ over a time period (Harris & Shen, 2006; Jorion, 2006). The expression of the value at risk is

$$VaR_\alpha = q_a,$$

where $q_a$ is the corresponding quantile of the return distribution. The criterion based on the value at risk is given by

$$HE_{VaR} = 1 - \frac{VaR_\alpha(\Delta V_H)}{VaR_\alpha(\Delta S)},$$

where $VaR_\alpha(\Delta V_H)$ and $VaR_\alpha(\Delta S)$ are the VaR of returns at confidence level $\alpha$ for the hedging portfolio and the spot contracts, respectively. In this study, we set the confidence level $\alpha$ at 5%.

### 2.3.2. Time series cross-validation method for hedging performance

This study constitutes a time series cross-validation method for model performance estimation. Most studies estimate hedging method's out-of-sample performance by randomly splitting the dataset into the training and testing sample sets by the timeline and using the hedge ratio fitted by the training sample set to estimate its performance in testing sample set. However, this process is questionable for several reasons (Bailey et al., 2015, 2016, 2018; Bailey and Prado, 2014). First, the historical data are only the realization of many stochastic routes, and, thus, only estimating the performance on one realized route is not sufficient. Second, for the negative relationship between the hedging horizon and the sample size, the performance estimation loses the effectiveness as the hedging horizon increases given the sample reduction problem. Third, the splitting is random for the training and testing samples, which can cause the non-robust estimate and data-snooping problem.

To solve these problems, this study constitutes a time series cross-validation method for hedging performance in spirit of the cross-validation method (Bailey and Lopez de Prado, 2012, 2014; Bailey et al., 2014, 2017), which is the original for supervised learning on financial data. The cross-validation method basically is a resample method for obtaining a robust estimate of model performance in supervised machine learning. The details of the methods are as follows: first, spilt the sample into $N$ groups and choose $k$ groups as testing set and other $N-k$ groups as training set, where according to the combination principle there are $\binom{N}{k}$ training–testing sets. Then, train the model using samples of the training set and estimate the model performance using samples of the testing set. Finally, average the model performance of all training–testing sets.

However, the cross-validation method in time series requires revision. In time series, the independent identical condition is not valid as in the conventional cross-validation method. Thus, the model out-performing over one specific period does not mean it out-performs in other periods. We need to estimate the model performance in the whole time series, which is named as path performance, instead of the testing set performance in the conventional cross-validation method. Another difference is that we pass a hedge ratio to the testing sets instead of the machine learning models, meaning that the performance model in testing sets is common, but the hedge ratios from different methods in training sets vary.

The details of the time series cross-validation method for hedging methods are as follows:

1. Divide the time series $T$ into $N$ groups without shuffling, where the sample size in $N-1$ groups is $\lfloor T/N \rfloor$, and the $Nth$ group contains the $T - \lfloor T/N \rfloor (N-1)$ residue samples.
2. Compute all possible training/testing splits, where for each spilt, $N-k$ groups constitute the training set and $k$ groups constitute the testing set.
3. Estimate the hedge ratio using each training set and estimate their hedging performance using the respective testing set. Finally, we obtain $\binom{N}{k}$ hedge ratios from the training sets and $\binom{N}{k}$ hedging performance estimation from the testing sets.
4. Compute the $p[N,k]$ performance paths, where $p[N,k] = \frac{k}{N}\binom{N}{k}$. As we have the performance estimation for each path, we can derive the empirical distribution of the method's performance.

### 2.3.3. Path performance

In Figure 1, suppose we set $N = 5$ and $k = 2$. Following the procedures above, there are 10 training-testing sets, 10 hedge ratios, 10 performance estimations, and 4 performance paths.

|    | S1 | S2 | S3 | S4 | S5 | S6 | S7 | S8 | S9 | S10 | Paths |
|----|----|----|----|----|----|----|----|----|----|-----|-------|
| G1 | 1  | 2  | 3  | 4  |    |    |    |    |    |     | 4     |
| G2 | 1  |    |    |    | 2  | 3  | 4  |    |    |     | 4     |
| G3 |    | 1  |    |    | 2  |    |    | 3  | 4  |     | 4     |

| | | | | | |
|---|---|---|---|---|---|
| G4 | 1 | 2 | 3 | 4 | 4 |
| G5 | | 1 | 2 | 3 | 4 | 4 |

**Figure 1. Combination of testing groups.** This figure presents the combination of testing groups of the time series cross-validation method for hedging method, and the parameters are $N = 5$ and $k = 2$. There are 10 training-testing sets denoted as S1–S10 and 5 groups denoted as G1–G5. The amount of paths that each group joins presents in the column "Paths". The numbers in the quadrant are the path number for every G-S cell.

The combination of testing groups is shown in Figure 1. Path 1 is the combination of $(G1, S1), (G2, S1), (G3, S2), (G4, S3), and (G5, S4)$ ; Path 2 is the combination of $(G1, S2), (G2, S5), (G3, S5), (G4, S6), and (G5, S7)$ ; Path 3 is combination of $(G1, S3), (G2, S6), (G3, S8), (G4, S8), and (G5, S9)$ ; Path 4 is combination of $(G1, S4), (G2, S7), (G3, S9), (G4, S10), and (G5, S10)$.

Thus, we can estimate the path performance of the hedging methods. The path performance estimators of hedging methods are extended from those of conventional performance models. For the variance reduction criterion $HE_{variance}$, we define the path variance reduction performance $HE_{variance}^{path}$ as

$$HE_{variance}^{path} = \frac{1}{N} \sum_{i}^{N} HE_{i,variance},$$

where $HE_{i,variance}$ is the $ith$ group's variance reduction performance and $N$ is the group number. For the VaR criterion, we define the path VaR performance $HE_{VaR}^{path}$ as

$$HE_{VaR}^{path} = \min(HE_{i,VaR}), i = 1, \ldots, N,$$

where $HE_{i,VaR}$ is the $ith$ group's VaR performance and N is the group number. The definition of path VaR performance is based on the idea of minmax principle that we take the performance in worst situation as the criterion, which is consistent with the VaR's object for estimating the extreme situation performance.

## 3. Empirical Results

The empirical results include the data and descriptive statistics, the preliminary analysis, the in-sample performance, the out-sample performance, and the determinants of the hedging performance. The out-sample performance and the determinants of the hedging performance are the core concerns for hedgers. We find the out-sample performance is inconsistent with the in-sample performance. The conventional method have superiority in performance on the variance reduction criterion, whereas the EMD method family has superiority in performance on the VaR criterion. The key determinant of the hedging performance is the matching degree of the spot and futures contracts at the corresponding hedging horizon, rather than the variance of the contracts.

## 3.1. Data and descriptive statistics

All data in this study are collected from DataStream. We test three main spot and futures contracts of commodity and financial index, including gold, Brent crude oil, and S&P 500 price index, because these contracts are the most representative in hedging management. We make the time series as long as possible to satisfy the sample size for more accurate analysis, so the period of sample is generally more than 10 years.

**Table 1. Data description.** This table presents the data information in this study. The contracts include the gold, Brent crude oil, and S&P500 price index. The futures contracts are the continuous near month contract corresponding to the spot contract. The period records the contract start date and end date. The sample size is the number of days in the period. The year is the number of years in the period. The exchanges of spot and futures contracts are in parentheses.

| Spot | Futures | Period | Sample Size | Year |
|---|---|---|---|---|
| Gold (Gold, Handy & Harman Base) | Gold (COMEX) | 1979.01.02–2021.10.29 | 11174 | 42 |
| Brent Crude Oil | Brent Crude Oil (ICE) | 2007.07.30–2021.10.29 | 3502 | 14 |
| S&P500 Price Index | Mini S&P500 Index (CME) | 1997.09.09–2021.10.29 | 6299 | 24 |

## 3.2. Preliminary Analysis

Before the in-sample and out-of-sample analysis, preliminary analyses of the IMF, including cycle, variance decomposition, and matching degree, are necessary.

### 3.2.1. Cycles of IMF

The IMFs except the residue or the trend have no trend but oscillation mode, from which the corresponding cycles can be calculated. Unlike the Fourier transform and wavelet method, the IMFs have time varying frequency because they contain the local information of the original time series, which makes calculating the constant frequency difficult. Following the conventions in Huang (1998, 2003), this study calculates the cycle of the IMF using

$$Cycle_i = \frac{length(IMF_i)}{\#maxima + \#minima - 1} \times 2,$$

where $length(IMF)$ is the length of $ith$ IMF time series, and the $\#maxima$ and $\#minima$ are the numbers of local maxima and local minima of $ith$ IMF, respectively.

Using the definition above, we calculate the cycles of the IMF of three contracts in Table 2. The spot and futures contracts have the similar cycles from $IMF_1$ to $IMF_5$ and have the different cycles from $IMF_6$ to the last IMFs as different contracts have respective numbers of IMFs. The contracts have similar periodic characteristics up to 2-3 months but have the less similar periodic characteristics from 3 months to larger scales.

**Table 2. Cycles of the IMF.** This table presents the cycles of the IMF of spot and futures contract time series. The data unit is the day. "Gold" denotes the gold contract, "Brent" denotes the Brent crude oil contract, and "S&P 500" denotes the S&P 500 price index. Each contract has different numbers of IMFs. The max number is 10 of gold, and the blank means no more intrinsic functions are extracted.

|  |  | IMF | | | | | | | | | |
|---|---|---|---|---|---|---|---|---|---|---|---|
|  | Contract | IMF1 | IMF2 | IMF3 | IMF4 | IMF5 | IMF6 | IMF7 | IMF8 | IMF9 | IMF10 |
| Gold | Spot | 3.03 | 6.23 | 12.70 | 26.32 | 56.43 | 134.63 | 328.65 | 1064.19 | 2793.5 | 11174 |
|  | Futures | 2.95 | 5.97 | 12.51 | 25.69 | 59.28 | 124.16 | 282.89 | 770.62 | 2793.5 | 11174 |
| Brent | Spot | 3.12 | 6.83 | 15.39 | 34.50 | 93.39 | 259.41 | 1000.57 | 3502 |  |  |
|  | Futures | 3.14 | 6.61 | 15.06 | 32.88 | 90.96 | 225.94 | 583.67 | 3502 |  |  |
| S&P 500 | Spot | 3.07 | 6.74 | 14.30 | 30.88 | 68.84 | 165.76 | 419.93 | 1399.78 | 6299 |  |
|  | Futures | 3.06 | 6.29 | 12.95 | 26.98 | 63.63 | 136.93 | 323.03 | 1145.27 | 6299 |  |

### 3.2.2. Decomposition of the log return time series variance

The EMD method can decompose the log return time series of contracts, of which the variance is decomposed at each time scales as well. From the decomposition, the scales, having more volatility, generally tend to be more inconsistent between the spot and futures contracts. This intuition is demonstrated in the decomposition of three contracts in Table 3. It is common for three contracts that the variance is consistent with the frequency of the scales, which means that the higher frequency are the scales, the larger variance the scales have. More than 50 percent variance contributes to the first time scale of each contract, and more than 80 percent variance contributes to the first three time scale of each contract. The spot and futures contracts have the similar variance decomposition patterns.

For the absolute value of variance, gold and the S&P price index have similar variance decomposition, whereas the variance of Brent crude oil is much larger than those of the other two contracts. From the analysis of the cycles and the matching degrees of the contracts, the difference of the variance decomposition is not related with the cycle pattern and the matching degree. Brent crude oil has the largest variance in each time scale, but the cycles and the matching degrees of spot and futures at each time scale have no apparent difference. Volatility is not the factor that drives the different behaviors between the spot and futures contracts.

**Table 3. Decomposition of the log return time series variance in IMF variances.** This table presents the variance of the IMF decomposed from the log return time series. The rows of "Spot" and "Futures" presents the absolute value and the rows of "%" presents the percentages of the IMF variance to the original log return series variance. The contracts include gold, Brent crude oil, and S&P price index whose denotation is consistent with the former table.

|  |  | IMF | | | | | | | | | |
|---|---|---|---|---|---|---|---|---|---|---|---|
|  |  | IMF1 | IMF2 | IMF3 | IMF4 | IMF5 | IMF6 | IMF7 | IMF8 | IMF9 | IMF10 |
| Gold | Spot | 9.178 | 2.790 | 2.024 | 0.898 | 0.574 | 0.505 | 0.145 | 0.036 | 0.056 | 0.036 |
|  | % | 56.47 | 17.17 | 12.46 | 5.53 | 3.53 | 3.11 | 0.89 | 0.22 | 0.35 | 0.22 |
|  | Futures | 8.253 | 2.857 | 1.719 | 1.124 | 0.657 | 0.455 | 0.172 | 0.053 | 0.023 | 0.025 |
|  | % | 53.79 | 18.62 | 11.20 | 7.33 | 4.28 | 2.96 | 1.12 | 0.35 | 0.15 | 0.16 |
| Brent | Spot | 37.001 | 13.489 | 8.240 | 3.647 | 1.521 | 1.011 | 1.541 | 0.280 |  |  |
|  | % | 55.19 | 20.12 | 12.29 | 5.44 | 2.27 | 1.51 | 2.30 | 0.418 |  |  |
|  | Futures | 37.654 | 13.007 | 7.037 | 4.405 | 1.541 | 1.345 | 0.731 | 0.527 |  |  |
|  |  | 56.69 | 19.58 | 10.60 | 6.63 | 2.32 | 2.03 | 1.10 | 0.793 |  |  |
| S&P500 | Spot | 8.988 | 3.148 | 1.827 | 0.920 | 0.701 | 0.456 | 0.107 | 0.029 | 0.030 |  |
|  | % | 55.34 | 19.38 | 11.25 | 5.66 | 4.31 | 2.81 | 0.66 | 0.18 | 0.18 |  |
|  | Futures | 9.498 | 3.291 | 2.086 | 0.849 | 0.464 | 0.282 | 0.102 | 0.031 | 0.020 |  |
|  | % | 57.07 | 19.78 | 12.53 | 5.10 | 2.79 | 1.69 | 0.61 | 0.19 | 0.12 |  |

### 3.2.3. Matching degree of the spot and futures contract

As the EMD method extracts the characteristics of the spot and futures contracts at different scales, the matching degree of the spot and the futures contract at the corresponding scales is important for hedge ratio and hedging performance estimation. The more the two contracts match at the corresponding scales, the better the hedging performance is inclined to appear. In Table 4, this study examines the matching degree of the spot and futures IMF through the R square of the OLS regression,

$$IMF_{i,t}^S = a + \beta IMF_{i,t}^F + e_{i,t},$$

where the denotations are the same with those in the former equation.

The matching degree of the gold contracts is the lowest on average, whereas the matching degrees of Brent crude oil and the S&P 500 price index are much better. The matching degree has a U shape relation with the intrinsic mode cycles, meaning that the matching degree is relatively high at the short and long cycles but are relatively low at the medium cycles.

**Table 4. Matching degree of the spot and futures IMF.** This table presents the matching degree of IMF between the spot and futures contracts. The model used to estimate the matching degree is the OLS regression model, $IMF_{i,t}^S = a + \beta IMF_{i,t}^F + e_{i,t}$, where $IMF_{i,t}^S, IMF_{i,t}^F$ are the IMFs of spot and futures, respectively. The rows of "$\beta$" are parameter $\beta$ estimated from the OLS regression model above, and the rows of "$R^2$" are the R-square estimated from the OLS regression model above. The contracts include gold, Brent crude oil, and S&P500 price index, and the denotations are consistent with the former table. The last IMF of each contract is the trend decomposed from the original time series.

|  |  | IMF | | | | | | | | | | |
|---|---|---|---|---|---|---|---|---|---|---|---|---|
|  |  | IMF1 | IMF2 | IMF3 | IMF4 | IMF5 | IMF6 | IMF7 | IMF8 | IMF9 | IMF10 | IMF11 |
| Gold | $\beta$ | 0.4808 | 0.5415 | 0.5856 | 0.5228 | 0.6636 | 0.5689 | 0.7166 | 0.5890 | 0.9723 | 1.0504 | 0.9266 |
|  | $R^2$ | 0.1968 | 0.2998 | 0.3493 | 0.3158 | 0.2419 | 0.3367 | 0.2873 | 0.1758 | 0.8463 | 0.9908 | 0.9968 |
| Brent | $\beta$ | 0.9303 | 0.9385 | 0.8697 | 0.8848 | 0.6702 | 1.0580 | 0.8876 | 0.6254 | 0.8321 |  |  |
|  | $R^2$ | 0.8858 | 0.8027 | 0.7638 | 0.6735 | 0.6131 | 0.6883 | 0.3857 | 0.5372 | 0.9191 |  |  |
| S&P500 | $\beta$ | 0.9584 | 0.8682 | 0.7496 | 0.6472 | 0.7226 | 0.9818 | 1.0044 | 0.8022 | 0.7491 | 0.8751 |  |
|  | $R^2$ | 0.8738 | 0.6424 | 0.4684 | 0.3943 | 0.4162 | 0.2698 | 0.3650 | 0.3402 | 0.8377 | 0.9945 |  |

### 3.2.4. Spot and futures IMFs

Figures 2~4 present the IMFs of the spot and futures contracts, including gold, Brent crude oil, and S&P500 price index, which provide an intuitive understanding of the EMD method. From the previous analysis, the cycles of spot and futures are matched up to $IMF_5$ at most and diverged in some subsequent IMFs. The matching degree of the IMFs is partly verified in Figures 2~4 as well. We cannot distinguish the matching degree of high frequency IMFs, but we can distinguish the difference of matching degree of low frequency IMFs.

The figures show that the gold IMFs are less matched than those of Brent crude oil and S&P 500 price index. The last 2 IMFs and the trend series, especially the latter series, are relatively well matched, indicating that the hedge ratio can converge to the naïve hedge ratio as the hedging horizon increases to infinite (Chen et al., 2004). The medium IMFs (greater than cycles of $IMF5$) presents difference between the spot and futures, meaning that they have different characteristics at the corresponding scales.

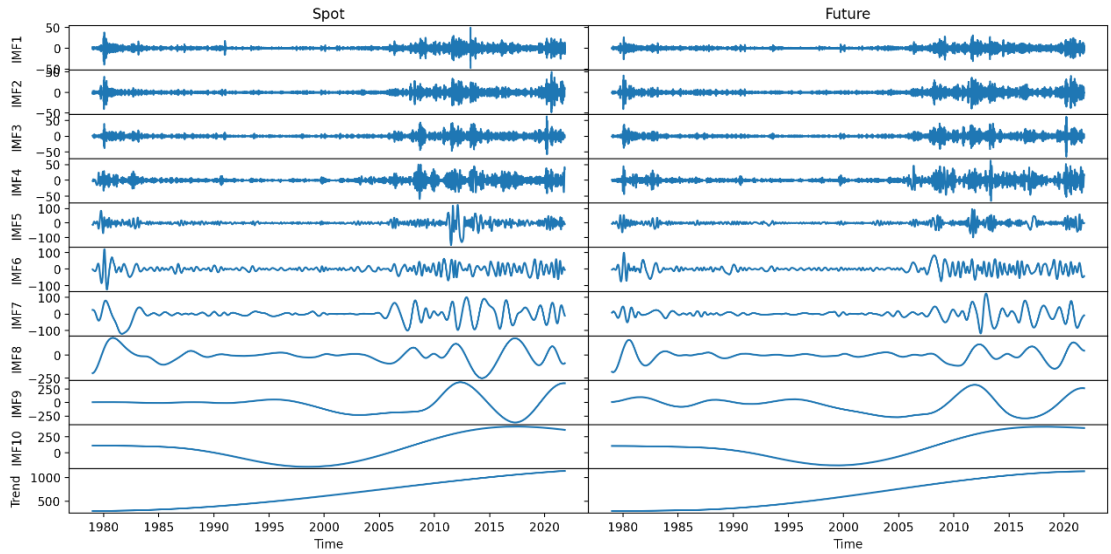

**Figure 2. Gold spot and futures IMFs.** This figure presents the IMF of the gold spot and futures contracts time series. The left plot with the title "Spot" is the spot IMF, and the right plot with the title "Future" is the futures IMFs. The transverse axis represents time period and the vertical axis represents the IMF and the trend.

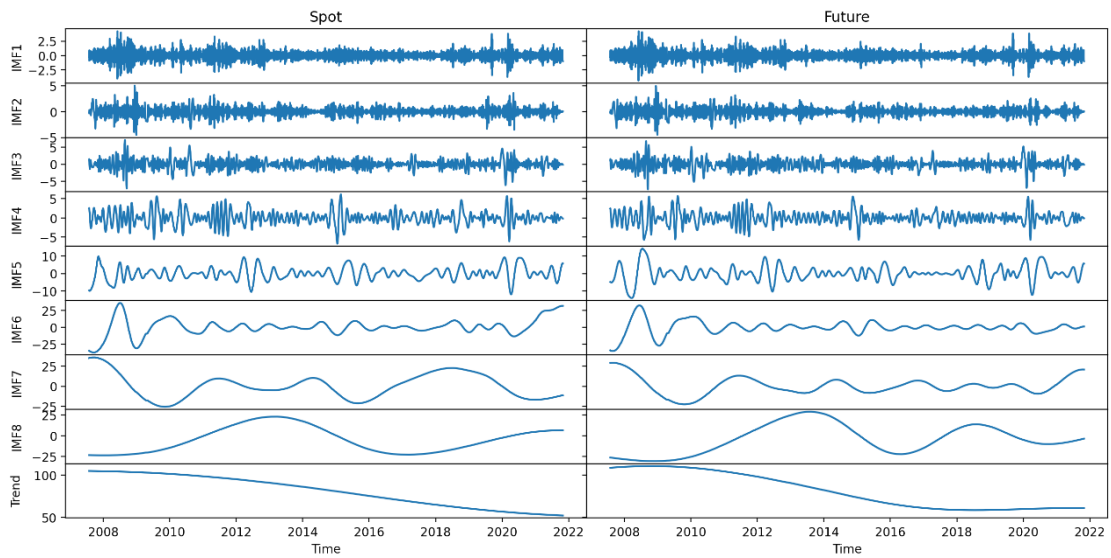

**Figure 3. Brent crude oil spot and futures IMFs.** This figure presents the IMF of the Brent crude oil spot and futures contracts time series. The left plot with the title "Spot" is the spot IMF, and the right plot with the title "Future" is the futures IMF. The transverse axis represents time period and the vertical axis represents the IMFs.

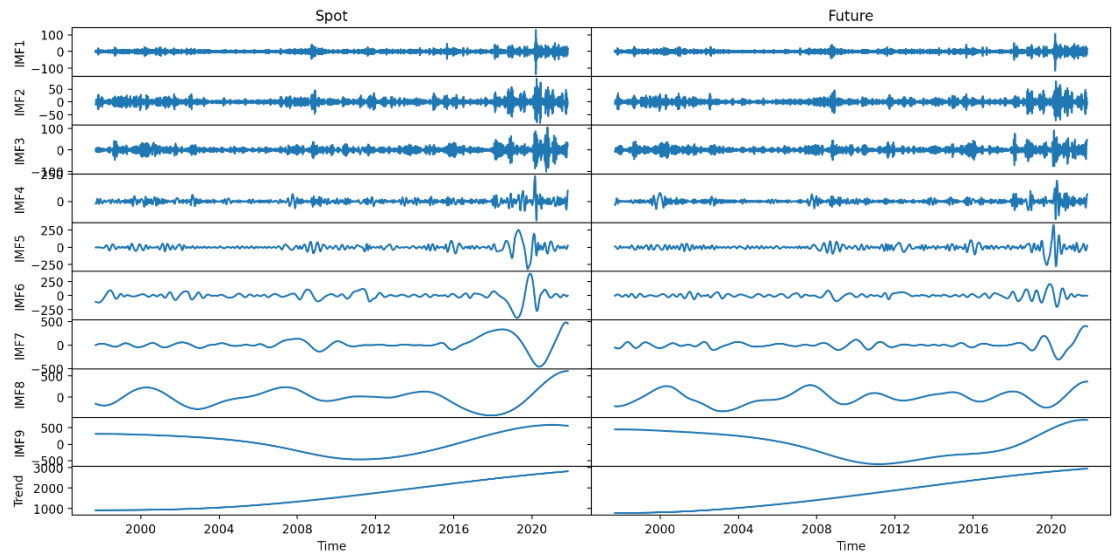

**Figure 4. S&P500 price index spot and futures IMFs.** This figure presents the IMF of the S&P500 price index spot and futures contracts time series. The left plot with the title "Spot" is the spot IMF, and the right plot with the title "Future" is the futures IMF. The transverse axis represents time period and the vertical axis represents the IMFs.

### 3.3. In-sample performance

We use conventional hedging methods, including minimum variance hedging method, error correction hedging method, and extended error correction hedging method, and the EMD-based hedging methods, including vanilla hedging method, sample saving hedging method, and aggregate hedging method, to estimate their in-sample hedging performance on variance reduction criterion and VaR criterion respectively. The in-sample performance estimation takes the whole time series as the training set and the testing set at the same time, so using the time series cross-validation method for hedging performance is insufficient. In addition, the in-sample performance is not the central concern of the hedgers. We define the cycles of the IMFs as the adaptive hedging horizons in the contracts.

For hedging performance on the variance reduction criterion in Table 5, same with Lien and Shrestha (2007),we find that at the short scales whose cycles are approximately in one month, the MV method shows significantly better performance than the EMD method family. At the long scales whose cycles are more than one month, the difference of performance between the MV method and the EMD family method tends to decrease dramatically. In most cases, the sample saving EMD method deteriorates the performance of the vanilla EMD method, which probably results from the different data processing methods. The performance models use the log return time series, whereas the EMD method family uses the original time series. In most cases, the aggregate EMD method improves the performance of the vanilla and sample saving EMD method, which is consistent with our former analysis. Unsurprisingly, the MV method provides the best in-sample variance reduction performance because its optimal function is identical to the performance criterion.

**Table 5. Hedge ratios and in-sample hedging performance (variance reduction).** This table presents the hedge ratios and in-sample hedging performance of variance reduction. The contracts include gold, Brent crude oil, and the S&P500 price index. The IMFs include those whose cycles are less than one year. The column "Cycle" is the rounding cycles of the corresponding IMFs. The hedging models used for calculating the hedge ratio include minimum variance hedging method (MV), error correction hedging method (ECM), extended error correction hedging method (EECM), vanilla EMD hedging method (VEMD), sample-saving EMD hedging method (SEMD), and aggregate EMD hedging method (AEMD), whose abbreviations are in the parentheses. The hedging performance model is the variance reduction criterion and the estimation values are expressed in percentage.

| | | Cycle | Hedge Ratio | | | | | | Hedge Effectiveness (VR) | | | | | |
|---|---|---|---|---|---|---|---|---|---|---|---|---|---|---|
| | | | MV | ECM | EECM | VEMD | SEMD | AEMD | MV | ECM | EECM | VEMD | SEMD | AEMD |
| Gold | IMF1 | 3 | 0.7901 | 0.7933 | 0.7984 | 0.5180 | 0.4808 | 0.4808 | 64.0221 | 64.0211 | 64.0151 | 56.4254 | 54.2086 | 54.2086 |
| | IMF2 | 6 | 0.8598 | 0.8829 | 0.8886 | 0.6555 | 0.5415 | 0.6289 | 79.3340 | 79.2770 | 79.2450 | 74.8516 | 68.4573 | 73.6127 |
| | IMF3 | 13 | 0.9623 | 0.9596 | 0.9609 | 0.6234 | 0.5856 | 0.7270 | 90.5265 | 90.5258 | 90.5263 | 79.2980 | 76.6504 | 85.1113 |
| | IMF4 | 26 | 0.9869 | 0.9826 | 0.9823 | 0.7021 | 0.5228 | 0.7516 | 94.8324 | 94.8306 | 94.8304 | 86.9332 | 73.8625 | 89.4388 |
| | IMF5 | 56 | 0.9633 | 0.9779 | 0.9740 | 0.9370 | 0.6636 | 0.8078 | 97.4257 | 97.4031 | 97.4136 | 97.3534 | 87.9990 | 94.8869 |
| | IMF6 | 135 | 0.9716 | 0.9860 | 0.9782 | 0.6858 | 0.5689 | 0.7915 | 98.9924 | 98.9709 | 98.9878 | 90.4272 | 81.9843 | 95.5895 |
| | IMF7 | 329 | 0.9797 | 0.9861 | 0.9872 | 0.7832 | 0.7166 | 0.8810 | 99.7150 | 99.7107 | 99.7091 | 95.7052 | 92.5236 | 98.7036 |
| Brent Oil | IMF1 | 3 | 0.9823 | 0.9823 | 0.9759 | 0.9370 | 0.9303 | 0.9303 | 95.1456 | 95.1456 | 95.1417 | 94.9436 | 94.8790 | 94.8790 |
| | IMF2 | 7 | 0.9695 | 0.9811 | 0.9799 | 0.9400 | 0.9385 | 0.9454 | 95.5795 | 95.5659 | 95.5685 | 95.4910 | 95.4815 | 95.5204 |
| | IMF3 | 15 | 1.0023 | 1.0012 | 1.0007 | 0.8677 | 0.8697 | 0.9173 | 98.8178 | 98.8176 | 98.8175 | 97.0341 | 97.0869 | 98.1074 |
| | IMF4 | 35 | 0.9865 | 0.9829 | 0.9821 | 0.9121 | 0.8848 | 0.9389 | 99.0214 | 99.0201 | 99.0194 | 98.4587 | 97.9689 | 98.7905 |
| | IMF5 | 93 | 1.0015 | 1.0025 | 1.0032 | 0.7286 | 0.6702 | 0.7879 | 99.9635 | 99.9634 | 99.9632 | 92.5418 | 89.0239 | 95.4170 |
| | IMF6 | 259 | 0.9967 | 0.9989 | 1.0093 | 0.9181 | 1.0580 | 1.0336 | 99.8570 | 99.8565 | 99.8411 | 99.2355 | 99.4797 | 99.7205 |
| S&P500 | IMF1 | 3 | 0.9684 | 0.9709 | 0.9736 | 0.9736 | 0.9584 | 0.9584 | 97.4703 | 97.4697 | 97.4675 | 97.4675 | 97.4600 | 97.4600 |
| | IMF2 | 7 | 0.9755 | 0.9777 | 0.9780 | 0.8449 | 0.8682 | 0.9404 | 98.6725 | 98.6720 | 98.6719 | 96.9044 | 97.4784 | 98.5451 |
| | IMF3 | 14 | 0.9836 | 0.9870 | 0.9894 | 0.8050 | 0.7496 | 0.9368 | 99.0176 | 99.0164 | 99.0141 | 95.7515 | 93.4124 | 98.7930 |
| | IMF4 | 31 | 0.9905 | 0.9931 | 0.9938 | 0.6097 | 0.6472 | 0.8614 | 99.4222 | 99.4215 | 99.4211 | 84.7280 | 87.4785 | 97.7324 |
| | IMF5 | 69 | 0.9827 | 0.9857 | 0.9895 | 0.8034 | 0.7226 | 0.9473 | 99.7188 | 99.7178 | 99.7141 | 96.3981 | 92.7322 | 99.5890 |
| | IMF6 | 166 | 0.9893 | 0.9905 | 0.9889 | 1.3910 | 0.9818 | 0.9540 | 99.9172 | 99.9171 | 99.9172 | 83.4453 | 99.9114 | 99.7899 |
| | IMF7 | 420 | 0.9824 | 0.9870 | 0.9817 | 0.8678 | 1.0044 | 1.0962 | 99.9626 | 99.9604 | 99.9625 | 98.6029 | 99.9124 | 98.6222 |

For hedging performance on the VaR criterion in Table 6, the results are quite complicated in that the conventional methods generally have better performance than the EMD method family. The vanilla EMD hedging method has generally better performance than the sample saving EMD hedging method. The aggregate EMD hedging method generally improves the sample saving EMD hedging method performance, but it occasionally improves the vanilla EMD hedging method performance.

**Table 6. In-sample hedging performance (VaR).** This table presents the in-sample hedging performance on the value-at-risk criterion. The contracts include gold, Brent crude oil, and the S&P500 price index. The IMFs include those whose cycles are less than one year. The column "Cycle" is the rounding cycles of the corresponding IMFs. The hedging performance model is the VaR criterion and the estimation values are expressed in percentage.

|  |  | Cycle | Hedge Effectiveness (VaR) | | | | | |
|---|---|---|---|---|---|---|---|---|
|  |  |  | MV | ECM | EECM | VEMD | SEMD | AEMD |
| Gold | IMF1 | 3 | 39.2824 | 39.1042 | 39.0630 | 32.8688 | 31.0705 | 31.0705 |
|  | IMF2 | 6 | 55.4993 | 56.4208 | 56.5121 | 50.0539 | 43.7554 | 49.0671 |
|  | IMF3 | 13 | 68.6380 | 68.7092 | 68.6758 | 52.1800 | 50.0175 | 59.0594 |
|  | IMF4 | 26 | 72.8089 | 72.7763 | 72.8234 | 62.8853 | 48.4568 | 67.5766 |
|  | IMF5 | 56 | 83.0478 | 83.7578 | 83.9450 | 83.6142 | 66.8291 | 78.6661 |
|  | IMF6 | 135 | 89.3674 | 90.1976 | 89.6787 | 67.0000 | 54.8789 | 78.2811 |
|  | IMF7 | 329 | 90.6208 | 90.8592 | 90.8997 | 77.1194 | 71.1066 | 86.5451 |
| Brent Oil | IMF1 | 3 | 96.6633 | 96.6706 | 95.6111 | 90.6061 | 89.7505 | 89.7505 |
|  | IMF2 | 7 | 91.3084 | 91.5543 | 91.5833 | 89.6809 | 89.5896 | 89.9072 |
|  | IMF3 | 15 | 99.6707 | 99.8322 | 99.8981 | 85.1096 | 85.2749 | 89.8871 |
|  | IMF4 | 35 | 96.4997 | 96.2877 | 96.2886 | 88.6183 | 86.0903 | 91.6575 |
|  | IMF5 | 93 | 99.8153 | 99.6845 | 99.6028 | 71.4080 | 65.6825 | 77.2200 |
|  | IMF6 | 259 | 98.2898 | 98.1831 | 97.6818 | 91.8083 | 94.8342 | 96.5113 |
| SP500 | IMF1 | 3 | 86.1325 | 86.2081 | 86.3997 | 86.3988 | 86.0297 | 86.0297 |
|  | IMF2 | 7 | 89.5633 | 89.8556 | 89.8304 | 82.6145 | 84.5918 | 88.8234 |
|  | IMF3 | 14 | 89.2702 | 89.1733 | 89.1048 | 81.6923 | 77.4480 | 89.2098 |
|  | IMF4 | 31 | 92.4683 | 92.4057 | 92.4134 | 61.3997 | 66.0455 | 86.1623 |
|  | IMF5 | 69 | 93.5840 | 93.4025 | 93.5044 | 83.0279 | 75.4919 | 92.8053 |
|  | IMF6 | 166 | 96.8711 | 97.0985 | 96.7927 | 51.1495 | 97.1294 | 96.4646 |
|  | IMF7 | 420 | 96.7212 | 96.3948 | 96.7751 | 87.4222 | 93.5403 | 75.5384 |

The in-sample hedging performance is not the central issue of hedging performance estimation because all hedgers use current hedge ratio to fulfill a hedge objective in the future. Thus, the out-sample performance is the key concern for hedgers. As the in-sample results frequently cannot extend to the out-sample results, the hedging performance is no exception.

### 3.4. Out-sample performance

The out-sample performance is the most important topic in the hedge decision because hedgers hedge primarily to achieve their hedge objective, such as minimize the volatility of the hedge portfolio and minimize the high dimensional risk. Given the hedge objective, hedgers must choose a hedge ratio to build the hedge portfolio and then to achieve the hedge objective in the future hedging horizon. As the hedging performance can only be achieved in the future, hedgers must determine the hedge portfolio at the present. Thus, the hedging model based on the in-sample performance or history data cannot assure their performance will continue in the future dataset, which is named out-sample performance because the time series are not stochastic samples like those in the classification problem. On the contrary, the out-sample performance of time series greatly differs from the in-sample performance, meaning that the model is wrong if the selection is based on the in-sample performance.

The out-sample performance of hedging method must be explored because few studies provide reasonable analysis. Most studies estimate the out-sample performance based on arbitrarily separating the time-series into two groups as the training group and the testing group, estimate the hedge ratio in the training group, and estimate the performance of the hedge ratio in the testing group. The separation are stochastic, so the out-sample performance of the model is relatively stochastic and data snooping may exist. As the historic time series is the one realization of the stochastic time series, the out-sample performance of the mode is less robust. Thus, we constitute a time series cross-validation method for the out-sample performance estimation, which can provide more robust and precise analysis. The details of the method are described in Section 2. Following the method, we divide the time series in years and take each year as a group. We choose several groups as testing groups and the remaining groups as training groups, in detail, that is 3 testing groups for gold contracts, 5 testing groups for Brent oil contracts, and 4 testing groups for S&P500 price index contracts. Some hedging horizons have inadequate samples in the specific group, which should be eliminated from the data. Finally, we construct the path performance for each contract and report the descriptive statistics of the performance of paths. We limit the adaptive hedging horizon to be lower than half year because the longer hedging horizons are not usually confronted.

For hedging performance on the variance reduction criterion in Table 7, conventional methods generally have better performance than the EMD hedging method family. The sample saving EMD hedging method has a similar performance to the vanilla EMD hedging method. However, in less matching degree contracts, the sample saving EMD hedging method deteriorates the performance of the vanilla EMD method. The aggregate EMD method can promote the performance of the vanilla and sample saving EMD hedging method in most cases.

For the contract performance, we find that gold generally has the lowest performance, which may result from the lowest matching degree of the spot and futures contracts. The brent crude oil and the S&P price index contracts have similar better performance, which is consistent with their better matching degree of the spot and the futures contracts. The performance difference between the conventional method and the EMD hedging method family is the highest in the gold contracts and is relatively small in the Brent crude oil and S&P price index contracts.

**Table 7. Out-sample hedging performance (variance reduction).** This table presents the out-sample hedging performance on variance reduction criterion of hedging methods. The contracts include gold, Brent crude oil, and the S&P500 price index. The column "Hori" presents the hedging horizon corresponding to the rounding cycles of the IMFs. The column "Path" presents the amount of path used for performance estimation. The hedging performance model is the path performance of variance reduction criterion.

|  | Hori | Path | Hedge Effectiveness : Variance Reduction ||||||||||||||
|---|---|---|---|---|---|---|---|---|---|---|---|---|---|---|---|---|---|
|  |  |  | MV |||| EMD |||| SEMD |||| AEMD ||||
|  |  |  | Mean | Std | Skew | Kurt | Mean | Std | Skew | Kurt | Mean | Std | Skew | Kurt | Mean | Std | Skew | Kurt |
| Gold | 3 | 861 | 0.7118 | 0.0002 | 1.2712 | 3.5615 | 0.6249 | 0.0056 | -.8728 | 1.7489 | 0.5901 | 0.0046 | -0.0198 | 1.5172 | 0.5901 | 0.0046 | -0.0198 | 1.5172 |
|  | 6 | 861 | 0.8362 | 0.0003 | 1.6977 | 4.3337 | 0.7940 | 0.0078 | -3.6195 | 13.3211 | 0.7053 | 0.0041 | -1.4036 | 3.8830 | 0.7623 | 0.0024 | -1.4892 | 3.0981 |
|  | 13 | 861 | 0.9137 | 0.0001 | -0.9987 | 2.3256 | 0.7893 | 0.0120 | 1.7870 | 4.9955 | 0.7803 | 0.0054 | -0.7634 | 3.9624 | 0.8642 | 0.0026 | -0.9144 | 2.0053 |
|  | 26 | 861 | 0.9513 | 0.0001 | -1.7406 | 4.0069 | 0.8936 | 0.0073 | -0.6366 | 4.2773 | 0.7484 | 0.0055 | -0.2809 | 0.8503 | 0.9024 | 0.0025 | -0.4782 | 1.2507 |
|  | 56 | 861 | 0.9613 | 0.0001 | -1.8589 | 6.2093 | 0.9567 | 0.0077 | -3.8135 | 19.6836 | 0.8601 | 0.0154 | -0.9250 | 3.9622 | 0.9318 | 0.0035 | 0.5812 | 1.8752 |
|  | 135 | 861 | 0.9543 | 0.0001 | -2.8818 | 10.7826 | 0.8944 | 0.0167 | -2.8930 | 10.4905 | 0.7893 | 0.0202 | 0.1717 | 2.6556 | 0.9210 | 0.0049 | 1.2159 | 3.6429 |
| Brent | 3 | 1001 | 0.9653 | 0.0001 | -0.4322 | 0.0283 | 0.9639 | 0.0008 | -1.3576 | 2.9763 | 0.9627 | 0.0012 | -1.3155 | 2.4568 | 0.9627 | 0.0012 | -1.3155 | 2.4568 |
|  | 7 | 1001 | 0.9834 | 0.0001 | -1.1384 | 1.5427 | 0.9820 | 0.0008 | -0.9193 | 0.9183 | 0.9809 | 0.0010 | -0.8399 | 0.9610 | 0.9816 | 0.0007 | -1.1040 | 1.8306 |
|  | 15 | 1001 | 0.9917 | 0.0001 | -1.9112 | 5.1188 | 0.9664 | 0.0065 | -0.5636 | 0.0259 | 0.9754 | 0.0049 | -0.8284 | 0.5649 | 0.9854 | 0.0023 | -0.8656 | 0.9441 |
|  | 35 | 1001 | 0.9955 | 0.0001 | -0.8401 | 0.9114 | 0.9748 | 0.0124 | -0.9587 | 0.8352 | 0.9808 | 0.0093 | -0.3648 | -0.2352 | 0.9920 | 0.0022 | -0.0357 | -0.6958 |
|  | 93 | 1001 | 0.9969 | 0.0001 | -3.0468 | 11.6486 | 0.8889 | 0.0712 | -0.1280 | -0.5839 | 0.8851 | 0.0612 | 0.3910 | -1.0256 | 0.9488 | 0.0273 | 0.6628 | -1.1285 |
| SP500 | 3 | 2024 | 0.9786 | 0.0001 | -0.3223 | 0.1968 | 0.9782 | 0.0004 | -1.8274 | 6.3841 | 0.9782 | 0.0005 | -2.4246 | 5.2967 | 0.9782 | 0.0005 | -2.4246 | 5.2967 |
|  | 7 | 2024 | 0.9861 | 0.0001 | -0.2697 | 0.4708 | 0.9680 | 0.0032 | 0.1825 | 0.2174 | 0.9734 | 0.0024 | -0.8571 | 2.1609 | 0.9845 | 0.0007 | -1.4989 | 3.7616 |
|  | 14 | 2024 | 0.9886 | 0.0001 | -0.2548 | 0.9574 | 0.9567 | 0.0044 | -0.8943 | 0.9294 | 0.9334 | 0.0079 | -0.9525 | 1.5113 | 0.9866 | 0.0007 | -1.7549 | 2.9497 |
|  | 31 | 2024 | 0.9906 | 0.0002 | 0.1111 | 0.4664 | 0.8067 | 0.0456 | 1.7053 | 2.0823 | 0.8807 | 0.0164 | 1.7434 | 2.5022 | 0.9775 | 0.0034 | 1.0796 | 0.6202 |

| | | | | | | | | | | | | | | | | | |
|---|---|---|---|---|---|---|---|---|---|---|---|---|---|---|---|---|---|
| | 69 | 1771 | 0.9964 | 0.0001 | -1.3095 | 3.3114 | 0.9795 | 0.0226 | -2.5284 | 5.3701 | 0.9173 | 0.0402 | -1.5262 | 2.3106 | 0.9936 | 0.0033 | -2.3320 | 3.8627 |
| | 166 | 1771 | 0.9889 | 0.0001 | 1.0426 | 0.9929 | 0.7983 | 0.0618 | 1.6229 | 1.8792 | 0.9766 | 0.0187 | -4.2296 | 26.5310 | 0.9874 | 0.0021 | -2.4953 | 4.8719 |

The hedging performance on the VaR criterion has quite different results with the in-sample performance. The in-sample performance of the conventional methods is better than that of the EMD hedging method family. This result is contrary to the out-sample performance.

For performance on the VaR criterion, we find that the EMD hedging method family performs generally better than the conventional methods. The sample saving EMD hedging method has no regular pattern with the vanilla EMD hedging method. The aggregate EMD hedging method generally improves the performance of the sample saving EMD hedging method and occasionally improves the performance of the vanilla EMD hedging method.

The determinants of the out-sample performance are apparently not related with the variance of the contracts. Brent crude oil has the largest variance, but its out-sample performance is not the worst. However, the matching degree of the spot and futures contracts seems to have relation with the out-sample performance, which we discuss in the subsequent section.

**Table 8. Out-sample hedging performance (VaR).** This table presents the out-sample hedging performance on value-at-risk criterion of hedging methods. The contracts include gold, Brent crude oil, and the S&P500 price index. The column "Hori" presents the hedging horizon corresponding to the rounding cycles of the IMFs. The column "Path" presents the amount of path used for hedging performance estimation. The hedging performance model is the path performance of VaR criterion.

| | Hori | Path | Hedge Effectiveness : Value at Risk | | | | | | | | | | | | | | | |
|---|---|---|---|---|---|---|---|---|---|---|---|---|---|---|---|---|---|---|
| | | | MV | | | | EMD | | | | SEMD | | | | AEMD | | | |
| | | | Mean | Std | Skew | Kurt | Mean | Std | Skew | Kurt | Mean | Std | Skew | Kurt | Mean | Std | Skew | Kurt |
| Gold | 3 | 861 | 0.1403 | 0.0043 | -0.1567 | -0.9256 | 0.1993 | 0.0053 | -2.9922 | 11.0696 | 0.1683 | 0.0032 | 2.0725 | 5.8405 | 0.1683 | 0.0032 | 2.0725 | 5.8405 |
| | 6 | 861 | 0.4270 | 0.0018 | -2.5941 | 12.7467 | 0.4047 | 0.0088 | -2.1569 | 5.8788 | 0.3448 | 0.0043 | -1.7972 | 2.8925 | 0.3742 | 0.0011 | -1.1428 | 1.8825 |
| | 13 | 861 | 0.3519 | 0.0006 | -1.2279 | 2.6437 | 0.4112 | 0.0109 | 0.1532 | 0.3724 | 0.4019 | 0.0075 | 1.0765 | 1.9113 | 0.5048 | 0.0032 | -0.5030 | 1.7423 |
| | 26 | 861 | 0.3204 | 0.0028 | 0.8230 | 9.7702 | 0.4715 | 0.0177 | -0.4734 | 2.5708 | 0.3229 | 0.0067 | 0.0807 | 0.9867 | 0.4893 | 0.0091 | 0.1272 | 1.3861 |
| | 56 | 861 | -0.4878 | 0.0691 | -2.2931 | 7.1359 | -0.3864 | 0.1805 | -0.1060 | 3.0062 | 0.2188 | 0.0708 | 1.3167 | 2.8012 | 0.0405 | 0.0393 | -1.3911 | 2.7258 |
| | 135 | 861 | -6.6416 | 0.4622 | -1.9842 | 8.4099 | -0.1986 | 0.1779 | -12.799 | 191.4475 | -0.1371 | 0.0359 | 1.5455 | 1.5490 | -0.6689 | 0.6675 | -3.3283 | 11.0017 |
| Brent | 3 | 1001 | 0.8614 | 0.0008 | 1.1917 | 0.8323 | 0.8561 | 0.0038 | -1.8577 | 8.1772 | 0.8529 | 0.0067 | -3.0291 | 14.0843 | 0.8529 | 0.0067 | -3.0291 | 14.0843 |
| | 7 | 1001 | 0.9079 | 0.0023 | 1.0834 | 0.7619 | 0.8966 | 0.0104 | -1.4391 | 1.9309 | 0.8850 | 0.0139 | -0.4432 | -0.7341 | 0.8954 | 0.0085 | -1.3545 | 2.9279 |
| | 15 | 1001 | 0.8052 | 0.0145 | -0.0528 | -0.6469 | 0.7821 | 0.0221 | 0.2008 | -0.2649 | 0.8206 | 0.0228 | -0.1871 | -0.3541 | 0.8710 | 0.0121 | -1.0265 | 0.9322 |
| | 35 | 1001 | 0.6528 | 0.0001 | 0.1445 | 0.7518 | 0.6532 | 0.1932 | -1.1713 | -0.1316 | 0.8015 | 0.0376 | -1.6230 | 5.2758 | 0.8328 | 0.0547 | -1.1403 | -0.1786 |
| | 93 | 1001 | 0.7944 | 0.0135 | 1.3186 | 3.6484 | 0.5255 | 0.1120 | 0.6545 | 0.6207 | 0.5582 | 0.0821 | 1.1358 | 1.4844 | 0.6527 | 0.0900 | 0.9868 | 0.0716 |

| | | | | | | | | | | | | | | | | | |
|---|---|---|---|---|---|---|---|---|---|---|---|---|---|---|---|---|---|
| SP500 | 3 | 2024 | 0.6857 | 0.0003 | 0.0569 | 0.0886 | 0.6848 | 0.0020 | 3.7683 | 20.6089 | 0.6854 | 0.0031 | 2.6365 | 6.0093 | 0.6854 | 0.0031 | 2.6365 | 6.0093 |
| | 7 | 2024 | 0.6982 | 0.0007 | 0.0280 | 2.0520 | 0.7277 | 0.0045 | 0.2770 | -1.1949 | 0.7264 | 0.0031 | 1.2160 | 0.4372 | 0.7130 | 0.0035 | 0.6993 | 0.6520 |
| | 14 | 2024 | 0.7387 | 0.0013 | 0.1051 | 1.0475 | 0.7500 | 0.0072 | -2.0739 | 4.7830 | 0.7054 | 0.0158 | -0.6712 | 1.1255 | 0.7513 | 0.0021 | 0.9714 | 0.9207 |
| | 31 | 2024 | 0.5832 | 0.0238 | -0.5826 | 0.1127 | 0.4847 | 0.0449 | 1.2543 | 2.1061 | 0.5773 | 0.0233 | 1.8095 | 3.1399 | 0.7589 | 0.0130 | 0.7992 | -0.1859 |
| | 69 | 2024 | 0.4147 | 0.0328 | 0.1622 | 0.0622 | 0.5585 | 0.4546 | -2.1556 | 2.9390 | -0.0775 | 0.2914 | 0.5104 | 1.4491 | 0.6666 | 0.2080 | -2.9839 | 10.0188 |
| | 166 | 1771 | 0.1487 | 0.1635 | 2.0635 | 3.2774 | 0.3125 | 0.5592 | 1.7443 | 2.2201 | -0.7950 | 3.5156 | -2.3991 | 4.0689 | 0.7531 | 0.0196 | -5.4426 | 97.6465 |

### 3.5. Determinants of the hedging performance

Intuitively, there exists a potential relation between the out-sample performance and the matching degree of the contracts. The out-sample performance is consistent with the matching degree rather than the variance of the contracts. Thus, we explore the potential relation through the regression,

$$out\ sample\ performance_i = \alpha + \beta * matching\ degree_i + e_i,$$

where $\alpha$ is the constant, $\beta$ is the coefficient that measures the potential relation, the out-sample performance of contract $i$ is presented in Tables 7 and 8, and the matching degree is the $R^2$ presented in Table 4.

We present the potential relationship between the out-sample performance and the matching degree both on the variance reduction criterion and the VaR criterion in Table 9. We find that on the variance reduction criterion, the out-sample performance has a positive correlation with the matching degree. As the matching degree of the spot and futures contracts increases at the time scale, the out-sample performance improves in the hedging horizon corresponding to the specific time scale. Regarding the value-at-risk criterion, the out-sample performances of the vanilla and aggregate EMD hedging method have a positive correlation with the matching degree, but they have no significant correlation with the MV and the sample saving EMD method. The matching degree increases the out-sample performance of the vanilla and aggregate EMD hedging method.

**Table 9. Relationship between the out-sample performance and matching degree.** This table presents the relation between the out-sample performance and the matching degree. Panel A presents the out-sample performance on the variance reduction criterion and Panel B presents the out-sample performance on the VaR criterion. $\beta$ is the regression coefficient, and the $R^2$ is the R square of the regression model, $out\ sample\ performance_i = \beta * matching\ degree_i + e_i$. $\alpha$ and $\beta_{affine}$ are the regression coefficients, and the $R^2$ is the R square of the regression model, $out\ sample\ performance_i = \alpha + \beta_{affine} * matching\ degree_i + e_i$. The numbers in parenthesis are t value and the ***, **, and * presents significance at 0.01, 0.05, and 0.10, respectively.

|  | MV | VEMD | SEMD | AEMD |
|---|---|---|---|---|
| Panel A | Variance Reduction | | | |
| $\beta$ | 1.6026*** | 1.5305*** | 1.5129*** | 1.5742*** |
|  | (9.707) | (10.690) | (11.241) | (10.168) |
| $R^2$ | 0.855 | 0.877 | 0.888 | 0.866 |
| $\alpha$ | 0.8746*** | 0.7493*** | 0.6974*** | 0.8053*** |
|  | (22.185) | (16.404) | (13.940) | (14.770) |
| $\beta_{affine}$ | 0.1548** | 0.2900*** | 0.3584*** | 0.2410** |
|  | (2.164) | (3.498) | (3.948) | (2.435) |
| $R^2$ | 0.238 | 0.449 | 0.510 | 0.283 |
| Panel B | Value at Risk | | | |
| $\beta$ | 0.5610 | 0.9795*** | 0.9178*** | 1.0753*** |
|  | (0.732) | (9.441) | (6.330) | (7.537) |
| $R^2$ | 0.032 | 0.848 | 0.715 | 0.780 |
| $\alpha$ | -1.1879 | -0.0824 | -0.2740 | 0.0538 |
|  | (-1.166) | (-0.579) | (-1.454) | (0.272) |
| $\beta_{affine}$ | 2.5275 | 1.1160*** | 1.3714*** | 0.9863** |
|  | (1.367) | (4.316) | (4.009) | (2.750) |
| $R^2$ | 0.111 | 0.554 | 0.517 | 0.335 |

The analysis above suggests that on the variance reduction criterion, the optimal hedge ratio is from the MV model, whereas on the VaR criterion, the optimal choice is conditional. We analyze the model choice on the VaR criterion through the regression model,

$$relative\ performance_i = \alpha + \beta \times matching\ degree_i + e_i.$$

where $\alpha$ is the constant and $\beta$ is the regression coefficient. The relative performance is the model performance on VaR criterion relative to the MV method performance, and the matching degree is presented in Table 4. The relative performance is a ratio performance defined as

$$relative\ performance = \frac{model\ performance - MV\ performance}{|MV\ performance|},$$

where the MV performance is the MV hedging method performance and the absolute value operator is to avoiding negative performance of the MV hedging method.

In the above model, if the model performance is superior than the MV method, then the constant $\alpha$ is positive. $\beta$ presents the superior degree of the model as the matching degree increases, which means that if $\beta$ is positive, then the superiority extends as the matching degree increases. If $\beta$ is negative, then the superiority decreases as the matching degree increases. In Table 10, we find that the vanilla and aggregate EMD hedging methods have significant superiority. As the matching degree increases, the superiority decreases. By contrast, the sample saving EMD method presents no superiority. Thus, if the matching degree is relatively low, the optimal choice is the vanilla and aggregate EMD methods. If the matching degree is relatively high, the optimal choice is between the MV method and the EMD method family.

Table 10. Relationship between the model relative performance on the VaR criterion and the matching degree. This table presents the relation between the model relative performance on the VaR criterion and the matching degree. $\alpha$ is the constant, $\beta$ is the regression coefficient and $R^2$ is the R square of the regression model, $relative\ performance_i = \alpha + \beta \times matching\ degree + e_i$. The numbers in parenthesis are t value, and the ***, **, and * presents the significance at 0.01, 0.05 and 0.10, respectively.

|  | VEMD | SEMD | AEMD |
|---|---|---|---|
| $\alpha$ | 0.6258*** (3.191) | -0.8781 (-0.883) | 1.4212*** (2.618) |
| $\beta$ | -0.8779** (-2.467) | 1.1563 (0.641) | -1.8695* (-1.897) |
| $R^2$ | 0.289 | 0.027 | 0.194 |

## 4. Conclusion

This study constitutes an adaptive hedging method based on EMD and a time series cross-validation method for robust hedging performance estimation of hedging methods.

The conclusion is multi-dimensional. We extract the inherent mode of the spot and futures contracts. The variance clusters in the first three time scales of the contracts, which is not the key determinant of the hedging performance. The key determinant of the hedging performance is the matching degree of the spot and futures contracts corresponding to the hedging horizon. The in-sample performance is inconsistent with the out-sample performance, and the robust estimation of out-sample performance is important for hedging method selection. The MV hedging method shows superior out-sample performance on the variance reduction criterion. The EMD method family exhibits superior out-sample performance on the VaR criterion.

The conclusion can help hedgers optimize their hedging strategy because they have multiple hedging objectives. The conclusion provides market regulators with reference for more efficient futures market management, which can improve the hedging performance.